\documentclass[aps,prc,twocolumn,superscriptaddress]{revtex4-2}
\usepackage{CJK}
\usepackage{graphicx}
\usepackage{hyperref}
\usepackage{amssymb}
\usepackage{amsmath}
\usepackage[normalem]{ulem}
\usepackage{url}
\usepackage{color}

\begin{document}
\begin{CJK*} {UTF8}{} 

\title{Influence of the treatment of initialization and mean-field potential on the neutron to proton yield ratios}
\author{Junping Yang}
\affiliation{China Institute of Atomic Energy, Beijing 102413, China}
\affiliation{Department of Physics, Guangxi Normal University, Guilin, 541004, China}
\author{Yingxun Zhang}
\email{zhyx@ciae.ac.cn}
\affiliation{China Institute of Atomic Energy, Beijing 102413, China}
\affiliation{Guangxi Key Laboratory of Nuclear Physics and Technology, Guangxi Normal University, Guilin, 541004, China}
\author{Ning Wang}
\affiliation{Guangxi Key Laboratory of Nuclear Physics and Technology, Guangxi Normal University, Guilin, 541004, China}
\affiliation{Department of Physics, Guangxi Normal University, Guilin, 541004, China}
\author{Zhuxia Li}
\affiliation{China Institute of Atomic Energy, Beijing 102413, China}







\date{\today}

\begin{abstract}
In this work, we firstly investigate how to reproduce and how well one can reproduce the Woods-Saxon density distribution of initial nuclei in the framework of the improved quantum molecular dynamics model. Then, we propose a new treatment for the initialization of nuclei which is correlated with the nucleonic mean-field potential by using the same potential energy density functional. In the mean field potential, the three-body force term is accurately calculated. Based on the new version of the model, the influences of precise calculations of the three-body force term, the slope of symmetry energy, the neutron-proton effective mass splitting, and the width of the wave packet on heavy ion collision observables, such as the neutron to proton yield ratios for emitted free nucleons [$R(n/p)$] and for coalescence invariant nucleons [$R_{ci}(n/p)$] for $^{124}$Sn+$^{112}$Sn at the beam energy of 200 MeV per nucleon, are discussed. Our calculations show that the spectra of neutron to proton yield ratios [$R(n/p)$] can be used to probe the slope of symmetry energy ($L$) and the neutron-proton effective mass splitting. In detail, the $R(n/p)$ in the low kinetic energy region can be used to probe the slope of symmetry energy ($L$). With a given $L$, the inclination of $R(n/p)$ to kinetic energy ($E_k$) can be used to probe the effective mass splitting. In the case where the neutron-proton effective mass splitting is fixed, $R(n/p)$ at high kinetic energy can also be used to learn the symmetry energy at suprasaturation density.

\end{abstract}

\pacs{21.60.Jz, 21.65.Ef, 24.10.Lx, 25.70.-z}

\maketitle
\end{CJK*}

\section{Introduction}

The isospin asymmetric nuclear equation of state is fundamental for understanding the objectives of both nuclear physics, such as the properties of neutron-rich nuclei\cite{Moller12,Daniel14,Centel09,Brown2000,LWChen2010,Liu10,HJiang12,Wang13} and the mechanism of heavy ion collisions\cite{BALi97,MBTsang04,LWChen05,Baran05,Famiano06,Zhang08,Tsang09,BALi08,Tsang12,BALi14,QHWu15,YXZhang20}, and astrophysics, such as the properties of neutron star masses, radii, and tidal deformability\cite{Abbott17,Abbott18,Annala18,Fattoyev18,Abbott19,Malik19,NBZhang19,CYTsang19,MBTsang19}. However, the theoretical predictions present large uncertainties on the isospin asymmetric nuclear equation of state away from normal density, especially the density dependence of the symmetry energy. In the laboratory, low-intermediate energy heavy ion collisions (HICs) can provide the constraints of symmetry energy from subnormal density to twice the saturation density by comparing the data of HIC observables, such as neutron to proton yield ratios\cite{BALi97,Famiano06,Zhang08}, triton to He$^3$ yield ratios\cite{QingfengLi05,GCYong09}, isospin diffusion\cite{MBTsang04,LWChen05,YXZhang20}, neutron excess\cite{YZhang17}, $\pi^-/\pi^+$ ratios\cite{BaoanLi02,YYLiu21}, collective flow\cite{Russotto16,YJWang20}, and so on, with the transport model calculations.

Among those transport model calculations of isospin sensitive HIC observables, the initialization of nuclei plays an important role, as does the isospin dependent mean-field potential. For example, consideration of the neutron skin in the initialization can influence the prediction of $\pi^-/\pi^+$ ratios\cite{Hartnack18,GFWei14} in peripheral HICs. In these pioneer calculations, the slope of symmetry energy and the thickness of neutron skin were treated separately. However, the theoretical calculations show that the neutron skin thickness of a heavy nucleus, which is given by the difference between the root-mean-square radii of neutrons and protons, i.e., $\Delta r_{np}=\langle r^2\rangle^{1/2}_n-\langle r^2 \rangle^{1/2}_p$, is strongly correlated to the slope of symmetry energy $L$\cite{Centel09,Brown2000,LWChen2010,YXZhang20}. Thus, a consistent treatment of neutron skin in the initialization and isospin dependent mean-field potential in nucleon propagation is highly desired in the development of transport models to reduce the uncertainties of symmetry energy constraint caused by separately treating $L$ and $\Delta r_{np}$.


In the QMD type models, each nucleon is represented by a Gaussian wave packet,
\begin{equation}
\label{wfqmd}
\phi_i(\mathbf{r})=\frac{1}{(2\pi\sigma_r^2)^{3/4}}\exp{\{-\frac{[\mathbf{r}-\mathbf{r}_i(t)]^2}{4\sigma_r^2}+i\mathbf{r}_i\cdot \mathbf{p}_i(t)\}}.
\end{equation}
The initial nuclei are prepared by sampling he centroids of wave packets $\mathbf{r}_i(t=0)$ in a hard sphere\cite{Xujun16,UrQMD98,Aichelin20,Souza94,Niita95,Maruyama90}. As a result, the sampled density profile of the nucleus has a larger tail than the required density due to the finite width of the Gaussian wave packet. This is the main difficulty in reproducing the density distribution calculated by microscopic nuclear models. Even for the simple Woods-Saxon (WS) density distribution,  which is wildly used in transport model simulations, it is still hard to reproduce and has been discussed in the transport model comparison project\cite{Xujun16}. There were some efforts to improve the description of the density profiles in the initialization. For example, in the ultrarelativistic quantum molecular dynamics (UrQMD) model, the centroids of wave packets are sampled within the radius which is reduced by half a layer of nucleons from the original nuclear radius\cite{UrQMD98}. The sampled density profile is similar to a Woods-Saxon density distribution, but they have different diffuseness values. In the parton-hadron quantum molecular dynamics (PHQMD)\cite{Aichelin20} and T\"{u}bingen quantum molecular dynamics (TuQMD) models\cite{cozma18}, a small width of Gaussian wave packet is used to well reproduce the Woods-Saxon density distribution. Thus, an important theoretical question in the QMD type models is how to reproduce and how well one can reproduce the required density profile of the initial nucleus.

In this work, we investigate how to reproduce the initial density distribution of the nucleus by doing the inverse Weierstrass transformation, and then redo the treatment of the initialization and mean-field potential in the improved quantum molecular dynamics (ImQMD) model with the same Skyrme energy density functional. The symmetry energy and neutron-proton effective mass splitting effects on the isospin sensitive observables, such as neutron to proton yield ratios, for the reaction of $^{124}$Sn+$^{112}$Sn at the beam energy of 200 MeV/u are examined and discussed. The new calculations show that the spectra of neutron to proton yield ratios can be used to probe $L$ and the neutron-proton effective mass splitting. In the case where the neutron-proton effective mass splitting is fixed, neutron to proton yield ratios at kinetic energy greater than 100 MeV can be used to probe the symmetry energy at suprasaturation density. 
\section{Theoretical approaches}

For describing HICs, the ImQMD model is adopted. The treatments of nucleon-nucleon collision, Pauli blocking, etc. are as the same as in the previous version of ImQMD. More details about them can be found in the review paper~\cite{Zhang20FOP}. The improvements we made in this work are in the initialization and nucleonic mean-field potential.

\subsection{Nucleonic Mean-field potential}
In the ImQMD model~\cite{YXZhang20}, the Skyrme type nucleonic potential energy density without the spin-orbit term is used:

\begin{eqnarray}
\label{eq:edfimqmd}
u(\mathbf r)_{sky}=&&\frac{\alpha}{2}\frac{\rho^2}{\rho_0} +\frac{\beta}{\eta+1}\frac{\rho^{\eta+1}}{\rho_0^\eta}+\frac{g_{sur}}{2\rho_0 }(\nabla \rho)^2\\\nonumber
&&+\frac{g_{sur,iso}}{\rho_0}[\nabla(\rho_n-\rho_p)]^2\\\nonumber
&&+A_{sym}\frac{\rho^2}{\rho_0}\delta^2+B_{sym}\frac{\rho^{\eta+1}}{\rho_0^\eta}\delta^2+u_{md}.\\\nonumber
\end{eqnarray}
The $\alpha$ is the parameter related to the two-body term, $\beta$ and $\eta$ are related to the three-body term, and $A_{sym}$ and $B_{sym}$ are the coefficients in the symmetry potential and come from the two- and the three-body interaction terms. The density $\rho$ is obtained by integrating the Wigner phase space density $f_i(\mathbf{r},\mathbf{p})=\frac{1}{(\pi\hbar)^3}e^{-\frac{(\mathbf{r}-\mathbf{r}_i)}{2\sigma_r^2}-\frac{(\mathbf{p}-\mathbf{p}_i)^2}{2\sigma_p^2}}$ in momentum space for all nucleons, i.e.,
\begin{eqnarray}
\rho(\mathbf{r})=\sum_{i=1}^A \rho_i(\mathbf{r})&=&\sum_{i=1}^A \int f_i(\mathbf{r},\mathbf{p})d^3\mathbf{p}\\\nonumber
&=&\sum_{i=1}^{A}\frac{1}{(2\pi\sigma_r^2)^{3/2}}e^{-\frac{(\mathbf{r}-\mathbf{r}_i)^2}{2\sigma_r^2}},
\end{eqnarray}
and $\delta=(\rho_n-\rho_p)/(\rho_n+\rho_p)$. Here, $\mathbf{r}_i$ and $\mathbf{p}_i$ are centroids of distribution, and also the variational parameters of Gaussian single-particle wave functions~\cite{Zhang20FOP}.

The Skyrme-type momentum dependent energy density functional $u_{md}$ is written based on its interaction form $\delta (\mathbf r_1-\mathbf r_2 ) (\mathbf p_1-\mathbf p_2 )^2$~\cite{Skyrme56}. In the ImQMD model,
\begin{eqnarray}
\label{eq:mdimqmd}
&&u_{md}(\mathbf{r},\{\mathbf{p}_i-\mathbf{p}_j\})\\\nonumber
&=&C_0\sum_{ij}\int d^3pd^3p' f_i(\mathbf r,\mathbf p)f_j(\mathbf r,\mathbf p')(\mathbf p-\mathbf p')^2\\\nonumber
&&+D_0\sum_{ij\in n}\int d^3pd^3p'f_i(\mathbf r,\mathbf p) f_j(\mathbf r,\mathbf p')(\mathbf p-\mathbf p')^2 \\\nonumber
&&+D_0\sum_{ij\in p}\int d^3p d^3p' f_i(\mathbf r,\mathbf p)f_j(\mathbf r,\mathbf p')(\mathbf p-\mathbf p')^2.
\end{eqnarray}
$C_0$, $D_0$ are parameters related to the momentum dependent interaction.

The parameters in Eqs.(\ref{eq:edfimqmd}) and (\ref{eq:mdimqmd}) can be obtained from the standard Skyrme interaction parameters as in Refs.\cite{YXZhang06,YXZhang20}. The connection between the seven parameters $\alpha$, $\beta$, $\eta$, $A_{sym}$, $B_{sym}$, $C_0$, $D_0$ used in the ImQMD model and the seven nuclear matter parameters-the saturation density $\rho_0$, binding energy at saturation density $E_0$, incompressibility $K_0$, symmetry energy coefficient $S_0$, the slope of symmetry energy $L$, isocalar effective mass $m_s^*$, and isovector effective mass $ m_v^*$-are given in Ref.~\cite{YXZhang20}. Thus, one can alternatively use $\rho_0$, $E_0$, $K_0$, $S_0$, $L$, $m_s^*$, $ m_v^*$,  as input to study the influence of different nuclear matter parameters.

Based on recent constraints of nuclear matter parameters related to symmetry energy from nuclei to neutron stars\cite{YXZhang20}, we choose $K_0=240$ MeV, $S_0=30$ MeV, $m_v^\ast/m=0.7$, and $m_s^\ast/m=0.8$, which are also in the reasonable region for the Skyrme type force\cite{Dutra12}. The $g_{sur}$ and $g_{sur,iso}$ values are taken as 24.5 and -4.99 MeVfm$^{2}$, respectively. $L$ is varied from 30 to 110 MeV to analyze the influence of $L$ on the spectra of n/p ratios. These values are the default parameters we used, and are listed in Table~\ref{tab:NMPara}. In addition, the influence of $S_0$, $m_s^\ast$, and the neutron-proton effective mass splitting $\Delta m_{np}^\ast=m_n^\ast-m_p^\ast$ on the n/p ratios are also discussed. One point we want to mention is that $\Delta m_{np}^\ast=m_n^\ast-m_p^\ast$ is alternatively described by $f_I=(m/m_s^\ast-m/m_v^\ast)=\frac{1}{2\delta}(m/m_n^\ast-m/m_p^\ast)$ in the following discussions, because it can be analytically incorporated into the transport code and is independent of isospin asymmetry.
\begin{table}[htbp]
\caption{\label{tab:NMPara}%
Values of the nuclear matter parameter used in the ImQMD-L model. $\rho_0$ is in fm$^{-3}$, $E_{0}$, $K_{0}$, $S_{0}$, $L$ are in MeV, $g_{sur}$ and $g_{sur,iso}$ are in MeVfm$^2$. $f_I=\frac{m}{m_s^\ast}-\frac{m}{m_v^\ast}$=-0.178.}
\centering
\begin{tabular}{lcccccccc}
\hline
\hline
$\ K_{0}$ & $S_{0}$ & $E_{0}$ & $\rho_0$ & $m_v^\ast/m$ & $m_s^\ast/m$  & $g_{sur}$ & $g_{sur,iso}$ & $L$\\
\hline
240 & 30 & -16 & 0.16 & 0.7 & 0.8 & 24.5 & -4.99 & 30,50,70,90,110 \\
\hline
\hline
\end{tabular}
\end{table}

The nucleonic force acting on the $i$th nucleon is
\begin{equation}
F_{i}=\dot{\mathbf{p}}_i=-\frac{\partial U(\{\mathbf{r}_i,\mathbf{p}_i\})}{\partial \mathbf {r}_i}.
\end{equation}
In the above formula, the potential energy $U(\{\mathbf{r}_i,\mathbf{p}_i\})$ is obtained by integrating the potential energy density in coordinate space, i.e., $U(\{\mathbf{r}_i,\mathbf{p}_i\})=\int u(\mathbf r) d^3\mathbf r$, and it is a function of $\mathbf{r}_i$ and $\mathbf{p}_i$. Because the density has a Gaussian form, all the integrals in potential energy calculations can be done analytically except for the three-body related terms with $\eta\ne 2$. When $\eta\ne 2$, one has to solve it with an approximation or numerical method. Usually, the three-body related potential energy is approximately calculated as
\begin{eqnarray}
\label{u3f}
\frac{\beta}{\eta+1}\int \frac{\rho^\eta}{\rho_0^\eta}\rho d^3\mathbf{r}&=&\frac{\beta}{\eta+1}\sum_{i=1}^N <\frac{\rho^\eta}{\rho_0^\eta}>_i\\\nonumber
&\approx& \frac{\beta}{\eta+1} \sum_{i=1}^{N} <\frac{\rho}{\rho_0}>_i^\eta+ O(\rho)
\end{eqnarray}
by omitting the higher order term\cite{Feng11,Cozma13,Sujun11,Zhang20FOP} or adjusting the value of $\sigma_r$ in the three-body term\cite{Maruyama90,Aichelin20}. The three-body force acting on particle $i$ is calculated according to the followings expression:
\begin{equation}
\label{old3f}
\dot{\mathbf{p}}_i=-\frac{\partial U_3}{\partial \mathbf {r}_i}=-\frac{\beta}{\eta+1} \frac{\partial }{\partial \mathbf{r_i}} \sum_{j=1}^N   \frac{\langle\rho\rangle^\eta_j}{\rho_0^\eta}.
\end{equation}
In the above formula, $\langle \rho\rangle_j=\sum_k \frac{1}{(4\pi\sigma_r^2)^{3/2}}e^{-\frac{(\mathbf{r}_j-\mathbf{r}_k)^2}{4\sigma_r^2}}$.
For uniform matter, this approximation is good enough since the density fluctuation is zero. 

However, the density variance in intermediate energy HICs is strong and time dependent. Thus, the influence of the higher order term in Eq. (\ref{u3f}) should not be neglected. In this work we exactly calculate the three-body term by using the numerical quadrature method. Further, the force acting on particle $i$ related to the three-body term is calculated as
\begin{equation}
\label{cal3f}
\dot{\mathbf{p}}_i=-\frac{\partial U_3}{\partial \mathbf {r}_i}=-\beta\rho_0 \int \frac{\rho^\eta}{\rho_0^\eta}\frac{\rho_i}{\rho_0} \frac{\mathbf{r}-\mathbf{r}_i}{\sigma_r^2}d^3 \mathbf r.
\end{equation}
The integral in Eq. (\ref{cal3f}) is solved by using the 11-point Gauss-Legendre quadrature method, and it gives a stronger three-body force in ImQMD-L than in ImQMD. To distinguish with the previous version of the ImQMD model, we named it ImQMD-L (L, means the lattice method) in the following discussions.

\subsection{Initialization of nucleus in ImQMD-L}
In this section, we will introduce how to obtain the neutron and proton Woods-Saxon density distribution of the initial nucleus with the same Skyrme energy density functional as in the mean-field propagation in the ImQMD-L model. Then, we investigate the criteria for reproducing the Woods-Saxon density distribution with a Gaussian wave packet. Then, the influences of different widths of wave packet, which lead to the different shapes of Woods-Saxon density distributions, on the stability and binding energy of the initial nucleus are discussed. Finally, we describe the method of initialization used in the ImQMD-L model.

\subsubsection{Density distribution of the initial nucleus with restricted density variational method}
In the calculations, we take the density distribution as a Woods-Saxon density function with
\begin{equation}
\rho_i=\rho_{0i}\frac{1}{1+\exp(\frac{r-R_{i}}{a_i})}, i={n, p}.
\end{equation}
Here, $\rho_{0n}$, $\rho_{0p}$, $R_{p}$, $R_{n}$, $a_p$, and $a_n$ are the saturation density, radius, and diffuseness values of proton and neutron density distributions, and they are obtained by minimizing the total energy of the system given by,
\begin{equation}
\label{eq:edf}
E=\int \mathcal{H} dr=\int\{\frac{\hbar^2}{2m}[\tau_n(\mathbf{r})+\tau_p(\mathbf{r})]+u_{sky}+u_{coul}\}dr,
\end{equation}
under the condition of the conservation of particle number in the system. This method is named as the restricted density variational method (RDV) method~\cite{MLiu06}.

The same semiclassical expression of the Skyrme energy density functional as in ImQMD, i.e., $u_{sky}$, is applied. One should note, that $u_{md}$ is reduced to
\begin{equation}
u_{md}=\frac{C_0}{2\hbar^2}\rho\tau+\frac{D_0}{2\hbar^2}(\rho_n\tau_n+\rho_p\tau_p)
\end{equation}
in the calculations of the nucleus in its ground state. The kinetic energy density $\tau_i$ in the RDV method is given by
\begin{eqnarray}
\label{eq:tau}
\tau_i(\mathbf{r})&=&\frac{3}{5}(3\pi^2)^{2/3}\rho_i^{5/3}+\frac{1}{36}\frac{(\nabla\rho_i)^2}{\rho_i}+\frac{1}{3}\triangle\rho_i\\\nonumber
&&+\frac{1}{6}\frac{\nabla\rho_i\nabla f_i+\rho_i\triangle f_i}{f_i}-\frac{1}{12}\rho_i(\frac{\nabla f_i}{f_i})^2\\\nonumber
&&+\frac{1}{2}\rho_i(\frac{2m}{\hbar^2}\frac{W_0}{2}\frac{\nabla(\rho+\rho_i)}{f_i})^2,
\end{eqnarray}
where we use the extended Thomas-Fermi (ETF) approach including all terms up to second order (ETF2) and fourth-order (ETF4) as in Ref.~\cite{Brack85}. $\rho_i$ denotes the proton and neutron densities of the nucleus, and $\rho=\rho_n+\rho_p$. $W_0$ is the strength of the spin-orbit interaction, and we set it to zero in order to use the same form of Skyrme energy density functional as in ImQMD-L; the parameter $f_i(\mathbf{r})$ is the same as in Ref.~\cite{MLiu06}. The calculated results of $a_{p}$, $R_{p}$, $a_{n}$, $R_{n}$, binding energy $B$, and rms radii for neutron and proton obtained by RDV with the Skyrme energy density functional are listed in Table~\ref{tab:aRB}.

\begin{table}[htbp]
\caption{\label{tab:aRB}%
 $a_{p}$, $R_{p}$, $a_{n}$, $R_{n}$, binding energy $B$, and rms radius for neutron and proton for $^{124}$Sn obtained with the RDV method. $L$ and $B$ are in MeV, $a_{p}$, $R_{p}$, $a_{n}$, $R_{n}$, ${<r_p^2>}^{1/2}$, and ${<r_n^2>}^{1/2}$ are in fm. The values in the bracket are the results obtained with given $a_{p}=a_n=f^{-1}(\sigma_r=1.287)$, for details, see Secs.~\ref{initWT} and ~\ref{initqmd}.}
\centering
\begin{tabular}{ccccccccccc}
\hline
\hline
$L$ & $a_{p}(a')$ & $R_{p}$ & $a_{n}(a')$  & $R_{n}$ & $B(B^*)$ & ${<r_p^2>}^{1/2}$&${<r_n^2>}^{1/2}$\\
\hline
30  & 0.414 & 5.733 & 0.514 & 5.777 &  -7.971 & 4.700 & 4.865   \\
  &  (0.743) &  & (0.743) &  & (-7.306) &(5.230) & (5.259)   \\
50  & 0.415  & 5.729 & 0.507 & 5.811 &  -8.021 & 4.698 & 4.881   \\
  &  (0.743) &  & (0.743) &  & (-7.357) & (5.227) & (5.281)  \\
70  &  0.419 & 5.707 & 0.503  & 5.838 & -8.073 & 4.687 & 4.893  \\
  &  (0.743) &  & (0.743) &  & (-7.420) & (5.213) & (5.299)  \\
90  &  0.422 & 5.686 & 0.496 & 5.872 & -8.129  & 4.676 & 4.907 \\
  &  (0.743) &  & (0.743) &  & (-7.479) & (5.199) & (5.321)  \\
110 &  0.426 & 5.656 & 0.487 & 5.909 & -8.191 & 4.659 & 4.922  \\
 & (0.743) &  & (0.743) &  & (-7.539) &  (5.179) & (5.346)  \\
\hline
\hline
\end{tabular}
\end{table}

\subsubsection{Criteria for reproducing the Woods-Saxon density distribution}
\label{initWT}
Following the wave function of a nucleon used in the quantum molecular dynamics model, the nuclear density can be written as,
\begin{equation}
\label{rho}
\rho (\mathbf{r})=\sum_i^A \rho_i (\mathbf{r}),
\end{equation}
where
\begin{equation}
\label{rhoi}
\rho_i(\mathbf{r})=\frac{1}{(2\pi\sigma_r^2)^{3/2}}\exp\{-\frac{(\mathbf{r}-\mathbf{r}_i)^2}{2\sigma_r^2}\}.
\end{equation}
To reproduce the Woods-Saxon density profile of a nucleus, such as $\rho^{ws}(\mathbf{r})$,
\begin{equation}
\label{WS}
\rho^{ws}(\mathbf{r})=\rho_0\frac{1}{1+\exp(\frac{|r|-R}{a})},
\end{equation}
where $\rho_0$ is the saturation density, $R$ is the half density radius, and $a$ is the diffuseness parameter, by the method of Monte-Carlo sampling, one has to know the distribution of the centroid of the wave packet, $\tilde{\rho}(\mathbf r_i)$, which satisfies
\begin{equation}
\label{rho_req}
  \rho^{ws}(\mathbf{r})=\sum_i^A \int d^3\mathbf{r}_i \tilde{\rho}(\mathbf{r}_i) \frac{1}{(2\pi\sigma_r^2)^{3/2}}e^{-\frac{(\mathbf{r}-\mathbf{r}_i)^2}{2\sigma_r^2}}.
\end{equation}

For a spherical nucleus, the form of $\rho^{ws}(\mathbf r)$ has spherical symmetry, and thus the density distribution along the $x$ direction can be written as
\begin{eqnarray}
\label{rhox_req}
  \rho^{ws}(x)&=&\sum_i^A \int_{-\infty}^{\infty} dx _i \tilde{\rho}(x_i) \frac{1}{(2\pi\sigma_r^2)^{1/2}}e^{-\frac{(x-x_i)^2}{2\sigma_r^2}} \\\nonumber
  &=&\sum_i^A \int_{-\infty}^{\infty} dx_i \tilde{\rho}(x_i) G(x-x_i).
\end{eqnarray}
Equation~(\ref{rhox_req}) is a generalized Weierstrass transform\cite{WT} with a kernel $G(x-x_i)=\frac{1}{(2\pi\sigma_r^2)^{1/2}}\exp\{-\frac{(x-x_i)^2}{2\sigma_r^2}\}$, where $x_i$ is the centroid of a Gaussian kernel. Thus, the key point for the initialization of a nucleus with a Gaussian wave packet is to do an inverse Weierstrass transform to get the solution of $\tilde{\rho}(x_i)$ at a given Gaussian kernel $G(x-x_i)$.

Since the centroids of nucleons are sampled randomly within the density distribution of the nucleus, $\tilde \rho(x_i)$ can be thought as being the same for all nucleons. Thus, Eq.~(\ref{rhox_req}) can be simplified as
\begin{equation}
\label{rhox_req-new}
  \rho^{ws}(x)=A \int_{-\infty}^{\infty} dX \tilde{\rho}(X) G(x-X),
\end{equation}
with one-dimensional distribution $\tilde \rho(x_i)=\tilde \rho(X)$, and where $A$ is the number of nucleons. Principally, the distribution of the centroid of a Gaussian, i.e., $\tilde{\rho}(X)$, should be obtained by doing the inverse Weierstrass transformation. However, it cannot be obtained analytically. We approximate the solution by using the form of $\tilde \rho(X)=\rho_0\Theta(R^{WT}-X)$, where $\Theta(R^{WT}-X)$ is a Heaviside step function. By best fitting $\rho^{WT}(x)=\int \tilde \rho(X)G(x-X)dX$ to the required density profile $\rho^{ws}(x)$, the parameter $R^{WT}$ and the $\sigma_r$ in $G(x-X)$ are obtained.

\begin{figure}[htbp]
\centering
\includegraphics[angle=0,scale=0.35]{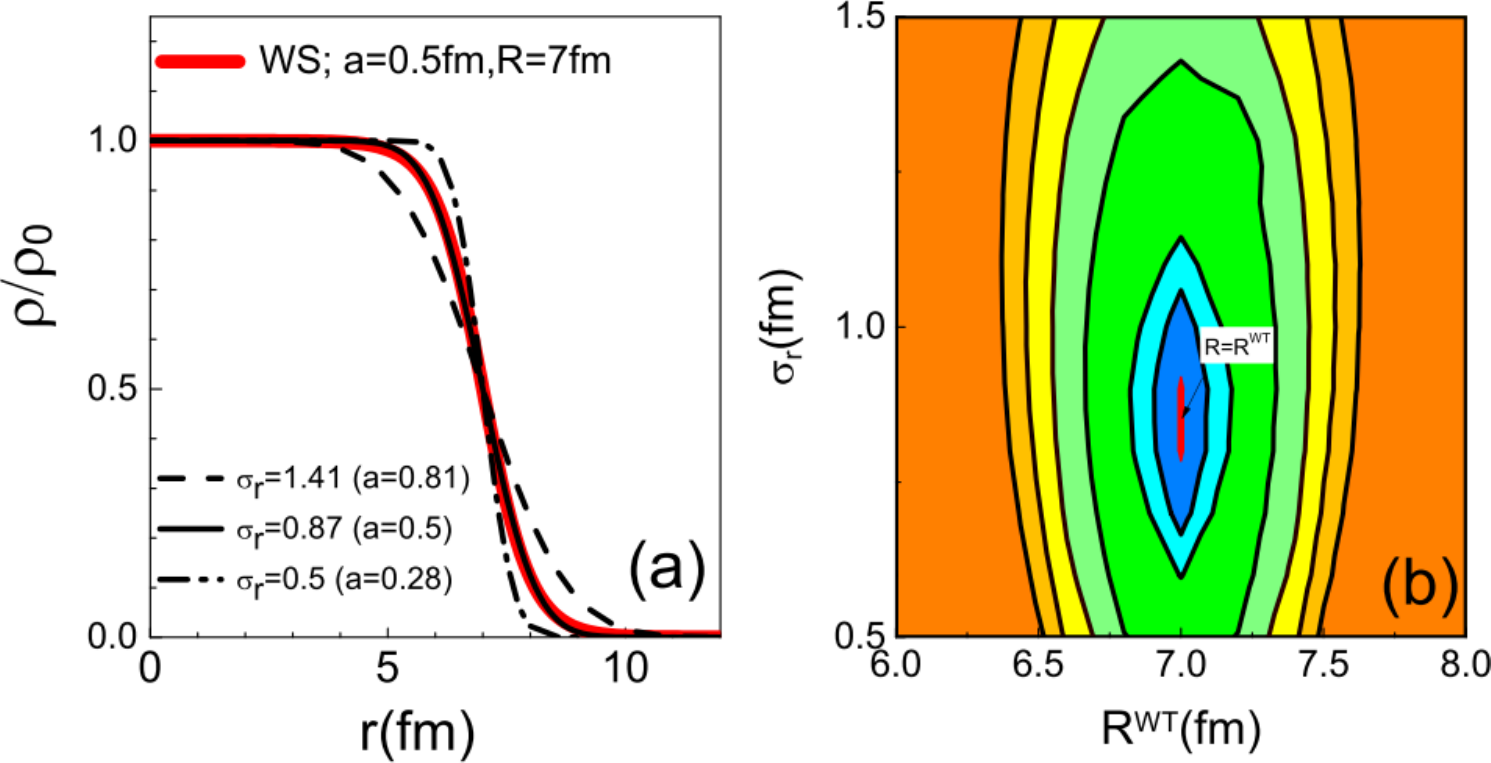}\\
\setlength{\abovecaptionskip}{0pt}
\vspace{0.5cm}
\caption{(Color online) (a) Woods-Saxon density distribution with $R$=7 fm and $a$=0.5 fm (red line), and Weierstrass transformation of $\Theta(R^{WT}-r)$ with different $\sigma_r$ at $R$=7 fm (black lines); (b) $\chi^2=\frac{1}{N}\sum_{i=1}^{N}{(\rho^{WT}_i-\rho^{WS}_i)}^2$ as a function of $\sigma_r$ and $R^{WT}$.}
\setlength{\belowcaptionskip}{0pt}
\label{fig1-a-R-chi2}
\end{figure}

In Fig.~\ref{fig1-a-R-chi2} (a), a Woods-Saxon density distribution with $R=7$ fm and $a=0.5$ fm is plotted as the red line. The black dashed line and dash-dotted line in panel (a) are the results obtained with different $\sigma_r$ at $R^{WT}=R=7$ fm. The results with $\sigma_r\approx 0.87$ fm are close to the Woods-Saxon distribution. Figure~\ref{fig1-a-R-chi2} (b) shows the $\chi^2$ of fitting $\rho^{WT}$ to $\rho^{ws}$ on $R^{WT}$ and the $\sigma_r$ plane. At $\chi^2=2.5\times 10^{-5}$, we obtain that $R^{WT}=R$, $\sigma_r=0.87$ fm. More generally, for reproducing the required Woods-Saxon distribution, two criteria should be fulfilled: a) the radius of the hard sphere is equal to half the density radius $R$, i.e., $R^{WT}=R$, and b)$\sigma_r$ is related to $a$ according to the relationship
\begin{equation}
\label{sigma-a}
 \sigma_r=f(a)=ka+c, R\ge 4.4a,
\end{equation}
where $k=1.71217\pm0.01548$ and $c=0.01564\pm0.01047$ fm should be fulfilled. For the Woods-Saxon distributions in three-dimensional space, the above criteria are also approximately available. 

\subsubsection{Stability of initial nucleus with different $\sigma_r$}
\label{initqmd}
In the QMD approach, the widths of the wave packets are time independent and is the same for the proton and neutron. Thus, one cannot exactly reproduce the density profile within the current framework of ImQMD due to the following points. First, the diffuseness parameters of the density distribution of sampled nuclei have $a_n=a_p$, which is not the same as in the distribution obtained with RDV where $a_n\ne a_p$. Second, the commonly used values of $\sigma_r$ in the QMD type models are larger than the values extracted based on Eq.~(\ref{sigma-a}). For example, $\sigma_r=1.414$ fm is usually used in the QMD calculations for Au+Au, but in the case of fitting the density distribution obtained by RDV the values of $\sigma_r$ for proton and neutron are $\sigma_r^p=$0.73 fm and $\sigma_r^n=0.88$ fm, respectively. It is found that the small $\sigma_r$ which can reproduce the density distribution obtained with RDV is worse in the stability of initial nuclei.

\begin{figure}[htbp]
\flushleft
\includegraphics[angle=0,scale=0.4]{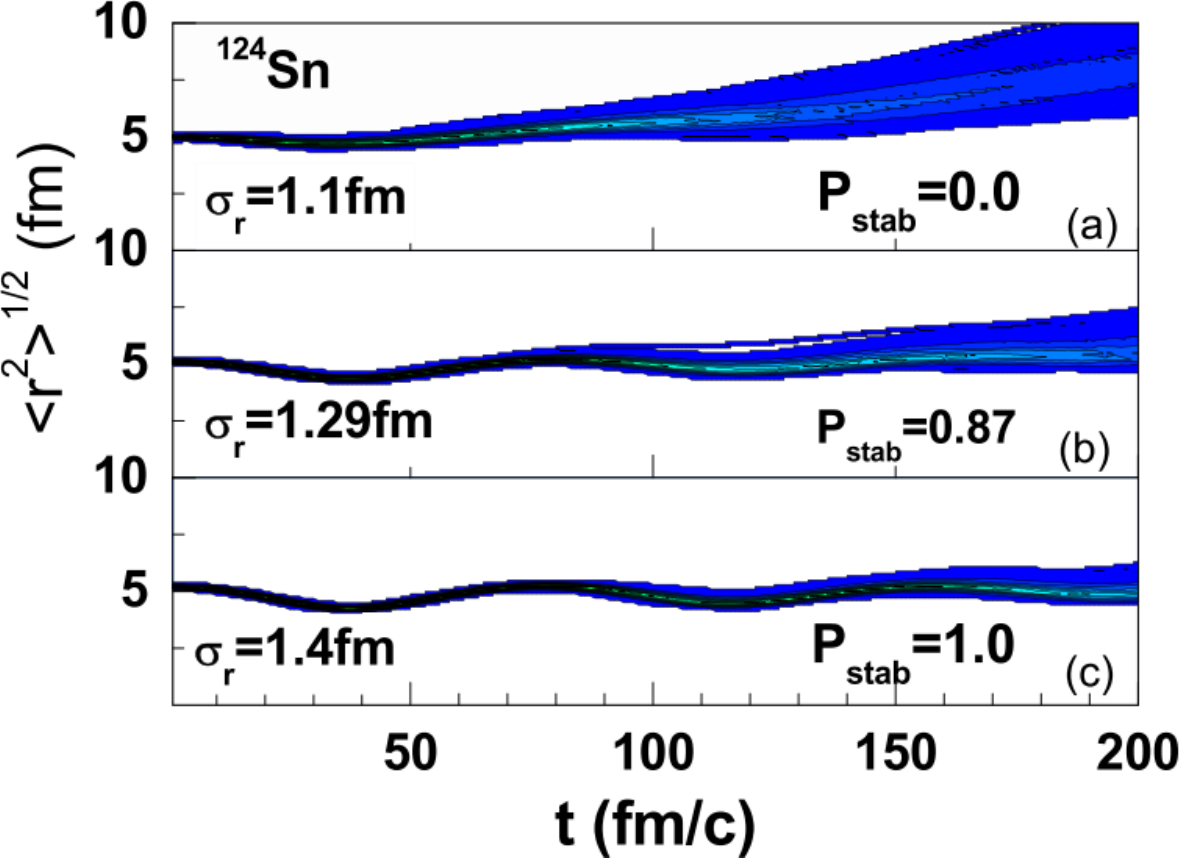}\\
\setlength{\abovecaptionskip}{0pt}
\vspace{0.2cm}
\caption{(Color online) Root mean square of $^{124}$Sn as a function of time. Panels (a), (b), and (c) are for $\sigma_r=1.1,1.29,1.4$fm, respectively. The results were obtained with 1000 events.}
\label{fig2-rms-t}
\end{figure}

To understand the influence of $\sigma_r$ on the stability of sampled nucleus, we present the root-mean-square ($rms$) values of sampled $^{124}$Sn in ImQMD-L as a function of time with three values of $\sigma_r$=1.1, 1.29, 1.4 fm in Fig.~\ref{fig2-rms-t}. In these calculations, the positions of neutrons and protons are sampled within $R_n$ and $R_p$, whose values are obtained by the RDV method with the same Skyrme energy density functional as that used in the mean-field propagation. To quantify the stability of the sampled initial nucleus, a variable named the probability of stability is used, and is defined as
\begin{equation}
  P_{stab}=\frac{N_{stab}}{N_{total}},
\end{equation}
where $N_{stab}$ is the number of events which keeps the $rms$ variation within 20\% at 200 fm/$c$ and $N_{tot}$ is the number of total events. For $\sigma_r$=1.1 fm, in panel (a), the $rms$ values obviously increase with time and all of them are larger than 20\% of $rms$ at t=0 fm/$c$, i.e., $P_{stab}=0.0$. 
The bad stability of the initial nucleus with smaller $\sigma_r$ is caused by a large initial fluctuation of the density distribution which may produce a stronger repulsive force on the particle. With the $\sigma_r$ increasing, the stability of the initial nucleus becomes better. $P_{stab}$ reaches about 87\% with $\sigma_r$=1.29 fm, and reaches about 100\% with $\sigma_r$=1.4 fm.

Consequently, the stability of the initial nucleus and reproducing the initial density profile cannot be achieved simultaneously in a QMD type model. To balance the accuracy of the reproduction of initial density distributions and stability of the initial nucleus, we select the width of the wave packet as $\sigma_r=1.287$ fm in the following $^{124}$Sn+$^{112}$Sn calculations.

\subsubsection{Initialization in ImQMD-L with $R_n$ and $R_p$}
In the initialization used in this work, the centroids of the wave packet for neutrons and protons are sampled within the half-density radii $R_n$ and $R_p$, and the binding energy of the sampled nucleus is in the range of $B\pm 0.2$ MeV. The values of $B$, $R_n$, and $R_p$ are calculated based on the RDV with the same Skyrme energy density functional as that used in the mean-field propagation of the ImQMD-L model. In Table~\ref{tab:aRB}, we present the $R_n$, $R_p$, and the binding energy $B$ of $^{124}$Sn obtained based on RDV with five parameter sets characterized by $L$=30, 50, 70, 90, 110 MeV, respectively. Thus, the neutron skin effect is correlated to the energy density functional used in the mean-field potential in ImQMD-L.  It is different from the method used in many other QMD codes, in which the $R_{n/p}$ is obtained by using the formula $r_0A^{1/3}$\cite{Xujun16}. 

For $^{124}$Sn+$^{112}$Sn, $\sigma_r=1.287$ fm is used and the sampled density distribution has a corresponding diffuseness with $a_n=a_p=a=f^{-1}(\sigma_r=1.287)=0.743$ fm. If there is no specification, $\sigma_r=1.287$ fm is the default value in the calculations. 
This $a$ value is larger than that obtained with the RDV method. Thus, the binding energy $B$ of the sampled nucleus deviates from the ground state energy. How much deviation of the binding energy is caused by the sampled density distribution with $a=0.743$ fm? In Table~\ref{tab:aRB}, the values in the brackets of the sixth column are the binding energies of $^{124}$Sn obtained with the RDV method for the Woods-Saxon density distribution with $R_n$, $R_p$, and $a=a_n=a_p=0.743$ fm. The binding energy per nucleon of $^{124}$Sn in the initialization with $a=0.743$ fm (or $\sigma_r=1.287$ fm) is about $\approx 0.65$ MeV larger than its ground state energy obtained with RDV, i.e., the binding energy deviation is about 8.2\%. It means that there is a spurious excitation in the QMD initialization.
ion.

\section{Results and discussions}
In the following studies, we perform the calculation of $^{124}Sn+^{112}Sn$ at the beam energy of 200 MeV/u, and impact parameter b=2 fm. Two isospin sensitive HIC observables will be discussed in this section. One is the neutron to proton yield ratio, i.e., $R(n/p)$, for emitted free nucleons,
\begin{equation}
R(n/p)=\frac{dY_{n}}{d E_k}/\frac{dY_{p}}{dE_k}.
\end{equation}
This observable is sensitive to the strength of the symmetry potential\cite{BALi97,Tsang09}, because the symmetry potential is opposite in sign for protons and neutrons. Another observable is the coalescence invariant neutron to proton yield ratio proposed in Ref.\cite{Tsang09,Famiano06}, which is defined as $R_{ci}(n/p)=\frac{dY^{CI}_{n}}{d(E_k/A)}/\frac{dY^{CI}_{p}}{d(E_k/A)}$, with
\begin{eqnarray}
\frac{dY^{CI}_{n}}{d(E_k/A)}&=&\sum_{A\le16,Z\le6} \frac{dY(Z,N)}{d(E_k/A)}\times N, \\\nonumber
\frac{dY^{CI}_{p}}{d(E_k/A)}&=&\sum_{A\le16,Z\le6} \frac{dY(Z,N)}{d(E_k/A)}\times Z.
\end{eqnarray}
The $R_{ci}(n/p)$ ratio still retain sensitivity to the symmetry energy, and it could eliminate the problem related to the absolute yield of light charged particles in the transport model simulations. Both $R(n/p)$ and $R_{ci}(n/p)$ are constructed from the transverse emitted nucleons and particles with an angular cut of $70^{\circ}\leq\theta_{c.m.}\leq110^{\circ}$.

In practical calculations, the exact values of $R(n/p)$ and $R_{ci}(n/p)$ also depend on the treatment of the three-body force in the mean-field potential. In this section, we reexamine the sensitivity of isospin sensitive observables to the slope of the symmetry energy ($L$) with the new treatment of the initialization and mean-field potential. In addition, the influences of the physical variables, such as $S_0$, $f_I$, $m_s^\ast$, and model parameter $\sigma_r$, are also discussed.


\subsection{Influences of the treatment of three-body force term and $L$ on the n/p ratios}

\begin{figure}[htbp]
\flushleft
\includegraphics[angle=0,scale=0.35]{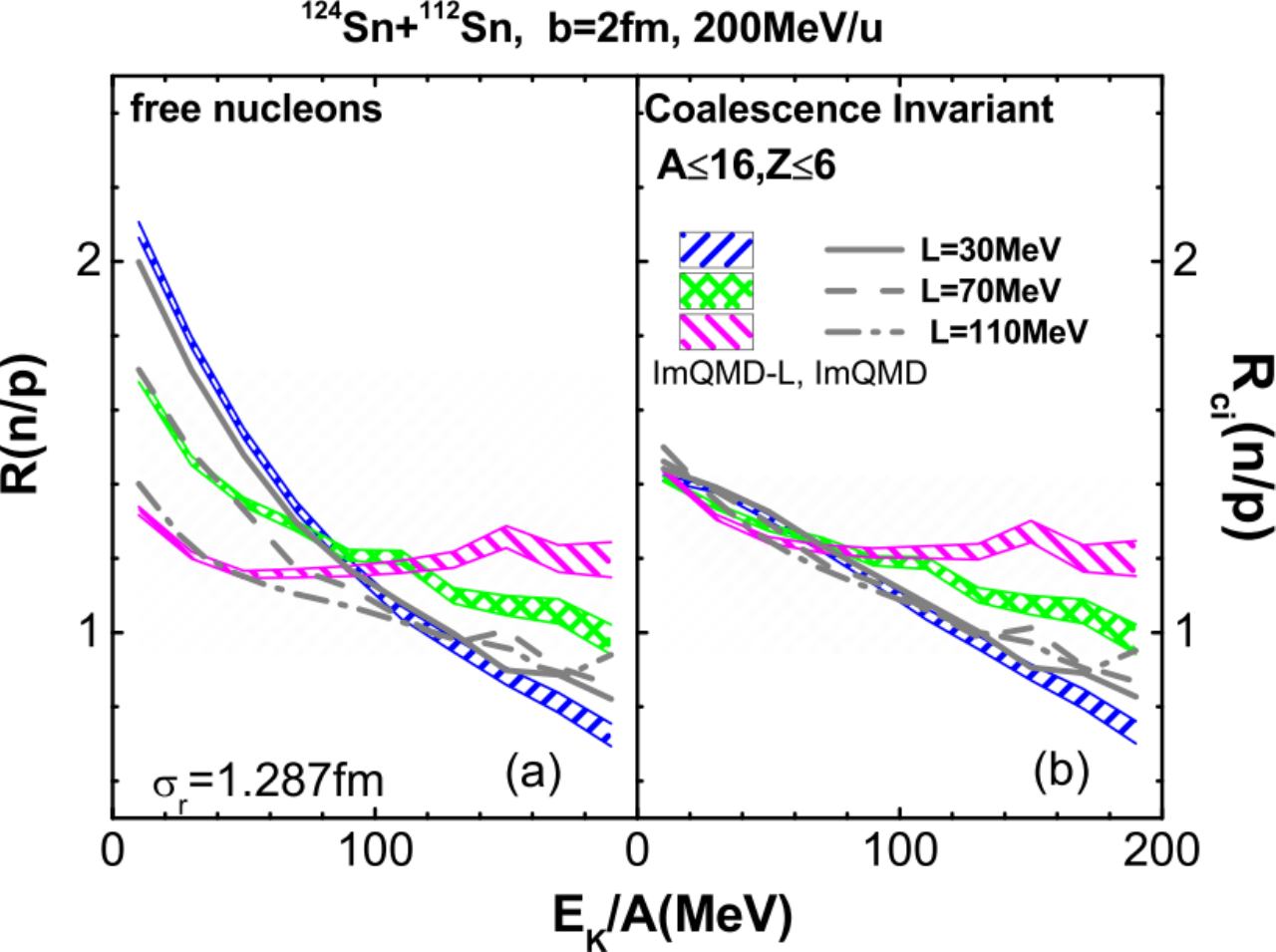}\\
\setlength{\abovecaptionskip}{0pt}
\vspace{0.2cm}
\caption{(Color online) Panels (a) and (b) are the neutron to proton yield ratios for emitted free nucleons and for coalescence invariant nucleons, respectively. Color shaded regions are the results from the new treatment of three-body force, and gray lines are the results from the old method (see text for more details).}
\setlength{\belowcaptionskip}{0pt}
\label{fig3-np-ek}
\end{figure}

Figure~\ref{fig3-np-ek} (a) shows the results of $R(n/p)$ obtained with ImQMD [gray lines, using Eq.(\ref{old3f}) to calculate the three-body force term], and ImQMD-L [color shaded region, using Eq.(\ref{cal3f}) to calculate the three-body force]. Three different symmetry energy cases, i.e., $L$=30 MeV (gray solid lines, blue shaded region), 70 MeV (gray dashed lines, green shaded region), 110 MeV (gray dash-dotted lines, red shaded region), are presented.

\setlength{\abovedisplayskip}{3pt}

As shown in panel (a), the $R(n/p)$ values of emitted free nucleons with $E_k<$80 MeV are greater for the soft symmetry energy case than for the stiff symmetry energy case, and $R(n/p)_{L=30}>R(n/p)_{L=70}>R(n/p)_{L=110}$. This is because the emitted nucleons with lower kinetic energy mainly come from the overlap region during the expansion phase. This region is below saturation density, where the symmetry energy is larger for smaller $L$. Larger symmetry energy in this region results in enhanced neutron emissions from the neutron-rich system and, thus, larger values of the $R(n/p)$ ratio. This behavior has been observed in other transport model predictions, especially for the widely studied Sn+Sn at the beam energy of 50 MeV/u\cite{Kong15}.

At $E_k>120$ MeV, there is $R(n/p)_{L=30}<R(n/p)_{L=70}<R(n/p)_{L=110}$. It is different than we observed at the beam energy around 50 MeV/u. This is because the emitted nucleons with high kinetic energy mainly come from the overlap region at the early stage of the expansion. During this stage, the density in the overlapped region is above the saturation density, where the stiff symmetry energy has larger values than the soft symmetry energy. It results in the $R(n/p)$ values being greater in the stiff symmetry energy case than in the soft symmetry energy case. The important finding is that the sensitivity of $R_{n/p}$ to $L$ becomes clearer within the calculations of ImQMD-L compared with the calculations of ImQMD. The reason is that the calculations in ImQMD-L provide stronger strength of the three-body force at high density, and thus of the strength of the symmetry potential.

In Fig.~\ref{fig3-np-ek} (b), the $R_{ci}(n/p)$ are presented. The sensitivity of $R_{ci}(n/p)$ to $L$ at low kinetic energy vanishes due to the contributions from the light particles and clusters. At high kinetic energy, the sensitivity of $R_{ci}(n/p)$ to $L$ is retained in the calculations with ImQMD-L.




\subsection{Influences of $S_0$, $f_I$ and $m_s^*$ on the n/p ratios}

As mentioned in Ref.\cite{YXZhang20}, the strength of the symmetry energy not only depends on $L$ but also depends on the symmetry energy coefficient $S_0$, isoscalar effective mass $m^\ast_s$, and $f_I$. However, the parameters $S_0$, $m_s^\ast$ and $f_I$ also have some uncertainties\cite{YXZhang20,Dutra12}, and it is worthwhile to understand their influence on $R(n/p)$. 

\begin{figure}[htbp]
\flushleft
\includegraphics[angle=0,scale=0.3]{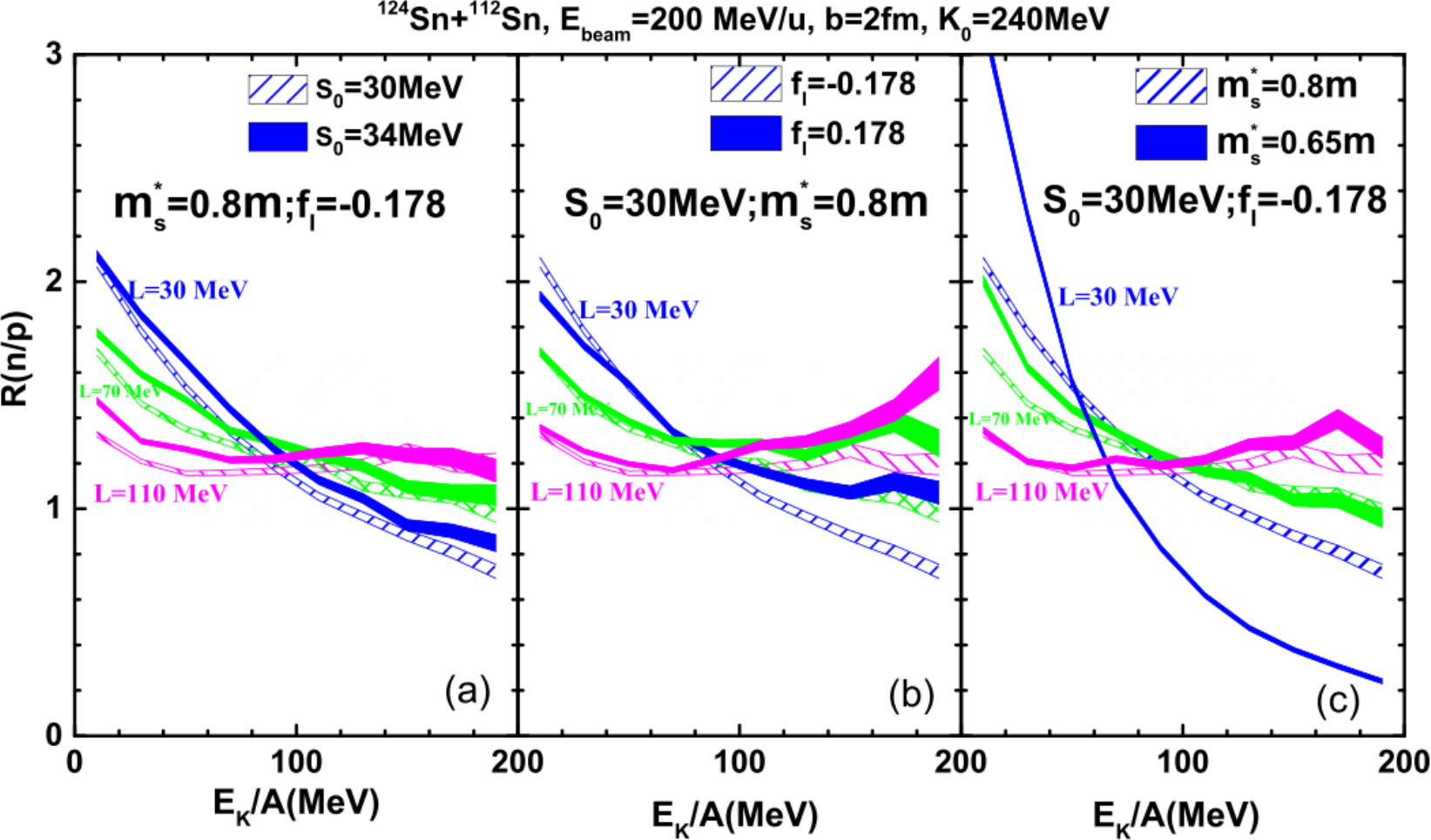}\\
\setlength{\abovecaptionskip}{0pt}
\vspace{0.2cm}
\caption{(Color online) The neutron to proton yield ratios for emitted free nucleons; different colors are for different $L$. Panel (a) is for different $S_0$, panel (b) is for different $f_I$, and panel (c) is for different $m_s^\ast$. }
\setlength{\belowcaptionskip}{0pt}
\label{fig4-np-s0fims}
\end{figure}

The results of $R(n/p)$ obtained with $S_0=34$ MeV are presented as solid shaded regions in panel (a) of Fig.\ref{fig4-np-s0fims}. All the other parameters are kept the same as in the default parameter sets. The different colors correspond to the results obtained with $L$=30, 70, 110 MeV. When a larger $S_0$ is adopted in the calculations, the curves of $R(n/p)$ shift up by $<$8\% owing to the enhanced symmetry energy in all density regions. However, the sensitivity of $R(n/p)$ to $L$ is not dramatically changed. 

\setlength{\abovedisplayskip}{3pt}

Another important ingredient that can influence $R(n/p)$ is the neutron-proton effective mass splitting, i.e., $m_n^\ast>m_p^\ast$ or $m_n^\ast<m_p^\ast$\cite{Zhang14}. In panel (b) of the Fig.\ref{fig4-np-s0fims}, the solid shaded regions are the results obtained with $m_n^\ast<m_p^\ast$, i.e., $f_I=0.178$, and different colors represent different $L$. For the given sets with $m_n^\ast<m_p^\ast$, i.e., $f_I=0.178$, the sensitivity of $R(n/p)$ to $L$ at lower kinetic energy region is similar to that obtained with default parameter sets where $f_I=-0.178$ ($m_n^\ast>m_p^\ast$). Thus, the $R(n/p)$ at $E_k\le 80$ MeV can help us to determine the values of $L$.

\begin{figure}[htbp]
\flushleft
\includegraphics[angle=0,scale=0.3]{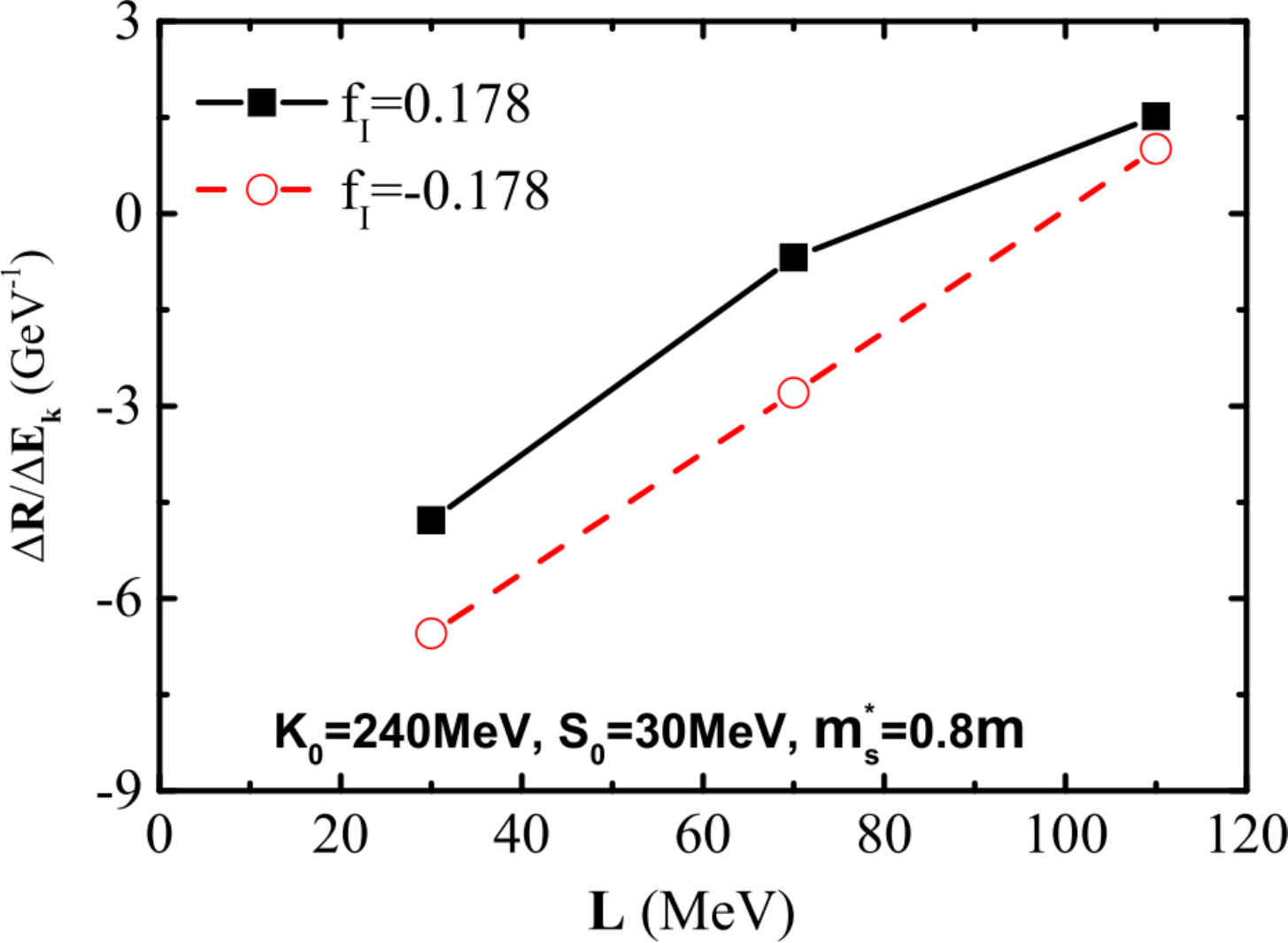}\\
\setlength{\abovecaptionskip}{0pt}
\vspace{0.2cm}
\caption{(Color online) The inclination of $R(n/p)$ to $E_k$ as a function of $L$ at $K_0$=240 MeV, $S_0$=30 MeV, and $m_s^*=0.8m$. Black solid symbols are for $f_I=0.178$, and open symbols are for $f_I=-0.178$.}
\setlength{\belowcaptionskip}{0pt}
\label{fig4-ms-fis}
\end{figure}

 In the high kinetic energy region, the sensitivity of $R(n/p)$ to $L$ becomes complicated if the neutron-proton effective mass splitting is not well fixed. For example, the values of $R(n/p)$ at $E_k> 100$ MeV obtained with $f_I$=0.178 and $L$=30 MeV are close to that obtained with $f_I$=-0.178 and $L$=70 MeV. This is because the sets with $f_I$=0.178 ($m_n^\ast<m_p^\ast$) have the strong lane potential as described in Ref.~\cite{Zhang14}, and it enhances the values of $R(n/p)$. To distinguish it, the inclination of $R(n/p)$ to $E_k$, which is calculated with the values of $R(n/p)$ obtained at $E^{(1)}_k$=50 MeV and $E^{(2)}_k$=150 MeV, i.e.,
\begin{equation}
RE_k=\frac{\Delta R(n/p)}{\Delta E_k},
\end{equation}
where $\Delta R(n/p)=R(n/p)(E^{(1)}_k)-R(n/p)(E^{(2)}_k)$ and $\Delta E_k=E^{(1)}_k-E^{(2)}_k$, can be adopted. The results of $\Delta R/\Delta E_k$ for different $f_I$ at $K_0$=240 MeV, $S_0$=30 MeV, and $m_s^*=0.8m$ are presented Fig.~\ref{fig4-ms-fis}, and the calculations with larger $f_I$ predict larger inclination.

Based on the above discussions, one can expect that the spectra of $R(n/p)$ can be used to obtain information on $L$ and $f_I$ simultaneously, if one does the analysis on the $L$ and $f_I$ parameter space.
\setlength{\abovedisplayskip}{3pt}

The influence of $m_s^\ast$ on $R(n/p)$ is also investigated in the case where other parameters are kept the same as in Table \ref{tab:NMPara}. The values of $R(n/p)$ obtained with $m_s^\ast=0.65m$ are presented as solid shaded region in panel (c) of Fig.~\ref{fig4-np-s0fims}. Different colors are the results obtained with different $L$. For $L$=70 and 110 MeV, reducing the isoscalar effective mass $m_s^\ast/m$ to 0.65 changes the $R(n/p)$ weakly. For $L$=30 MeV, obvious differences of $R(n/p)$ are observed between the set with $m_s^\ast=0.65m$ and the  default parameter set. The $R(n/p)$ obtained with $m_s^\ast/m=0.65$ are largely enhanced in the lower kinetic energy region and are obviously suppressed in the high kinetic energy region. This is because the set with $L$=30 MeV and $m_s^\ast/m=0.65$ have strong symmetry energy at low density and strong momentum dependent interaction, both providing a stronger repulsive force than other parameter sets and worsening the stability of the initial nucleus in the ImQMD-L simulations.

\subsection{Influence of $\sigma_r$ on the n/p ratios}
The width of wave packet $\sigma_r$ is an important model parameter in the QMD type models, and one can expect that the different values of $\sigma_r$ may influence the results. Thus, investigating the influence of $\sigma_r$ on the $R(n/p)$ within the ImQMD-L model can help us to understand the robustness of model calculations. In this work, the test calculations are performed with the value of $\sigma_r=1.0$ fm, which is not really used in low-intermediate energy HIC simulations.

\begin{figure}[htbp]
\center
\includegraphics[angle=0,scale=0.5]{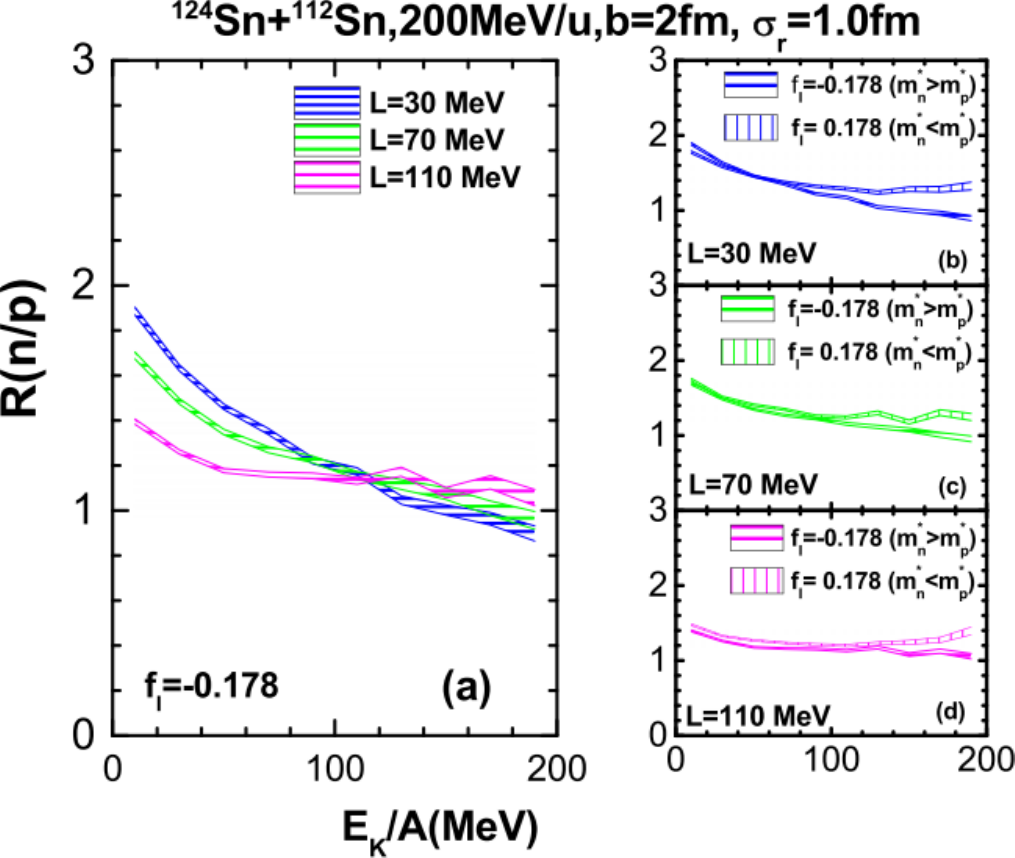}\\
\setlength{\abovecaptionskip}{0pt}
\vspace{0.2cm}
\caption{(Color online) Panels (a)-(d) are the neutron to proton yield ratios for emitted free nucleons obtained with different $L$ and different $f_I$ at $\sigma_r=1.0$ fm.}
\setlength{\belowcaptionskip}{0pt}
\label{fig3-np-sigmar}
\end{figure}

Figure~\ref{fig3-np-sigmar} (a) shows the $R(n/p)$ obtained with different $L$ at $\sigma_r=1.0$ fm. In the calculations with $\sigma_r=1.0$ fm, the values of $R(n/p)$ still depend on the slope of symmetry energy, and the behavior is similar to that with $\sigma_r=1.287$ fm. 
However, the sensitivity becomes weaker in both low and high kinetic energy regions than that with $\sigma_r=1.287$ fm. The reason is that the initial nuclei obtained with $\sigma_r=1.0$ fm in the ImQMD-L model have stronger initial fluctuation and worse stability than that with $\sigma_r=1.287$ fm. It increases spurious emission of nucleons with low kinetic energy. For example, for $L=30$ MeV, 22\% (18\%) more protons (neutrons) are emitted for the case of $\sigma_r=1.0$ fm than that with $\sigma_r=1.287$ fm. Consequently, the $R(n/p)$ obtained with $\sigma_r=1.0$ fm is smaller than that with $\sigma_r=1.287$ fm at lower kinetic energy. For $L=110$ MeV, 6\% (6.7\%) more protons (neutrons) are emitted for the case with $\sigma_r=1.0$ fm than that with $\sigma_r=1.287$ fm, and thus the $R(n/p)$ obtained with $\sigma_r=1.0$ fm is enhanced in the low kinetic energy region. Due to the conservation of nucleon number in the reaction system, $R(n/p)$ obtained with $\sigma_r=1.0$ fm is enhanced at high kinetic energy region for $L=30$ MeV, and suppressed for $L=110$ MeV. Thus, the effects of symmetry energy become relatively weak in the case of $\sigma_r=1.0$ fm. However, the sensitivity of $R(n/p)$ to $L$ is still large enough to distinguish the stiffness of symmetry energy.

\setlength{\abovedisplayskip}{3pt}

Figure~\ref{fig3-np-sigmar}(b), \ref{fig3-np-sigmar}(c) and \ref{fig3-np-sigmar}(d) show the results of $R(n/p)$ obtained with $f_I=-0.178$ ($m_n^*>m^*_p$) and $f_I=0.178$ ($m_n^*<m^*_p$) at $L$=30, 70, 110 MeV, respectively. As we discussed previously, the values of $R(n/p)$ in the high kinetic energy region obtained with $f_I=0.178$ ($m_n^*<m^*_p$) are greater than that obtained with $f_I=-0.178$ ($m_n^*>m^*_p$) at given $L$. The finding is similar to the results obtained with $\sigma_r=1.287$ fm, and it demonstrates that the small value of $\sigma_r=1$ fm do not dramatically change the sensitivity of $R(n/p)$ to the effective mass splitting.

\section{Summary and outlook}
In summary, we have investigated the issues of how to reproduce and how well to reproduce the Woods-Saxon density distribution with a Gaussian wave function. Based on the inverse Weierstrass transformation, we obtain two criteria for reproducing the Woods-Saxon density profile: one is that the centroid of the Gaussian wave packet should be sampled within a hard sphere with a radius approximately equal to the half density radius of the Woods-Saxon density distribution. Another is that the width of the Gaussian wave packet should be taken based on the parameter $a$, i.e. $\sigma_r= 0.01564+1.71217a$. The second condition requires a smaller width of Gaussian wave packet than the commonly used values in the quantum molecular dynamics model, and it causes initial nuclei to have bad stability due to the strong initial fluctuation. To balance the stability of initial nuclei and the requirement of initial density profile, the width of the Gaussian wave packet is taken as the one that can give the probability of stability greater than 85\% in this work.

To correlate the initialization and mean-field potential with the same energy density functional in the improved quantum molecular dynamics, we incorporate the RDV method into the ImQMD model. First, we calculate the half density radius of $R_n$, $R_p$, and the binding energy $B$ of the initial nuclei with the RDV method by using the same energy density functional as in the mean-field propagation. Then, the initial nuclei are sampled by using the obtained values of $R_n$, $R_p$, and binding energy $B$. In the mean-field part, we precisely calculate the three-body related term by using the 11-point Gauss-Legendre quadrature method at the expense of 50 times longer CPU times. These methods improve the theoretical reliability of the transport model and reduce the uncertainties of theoretical predictions owing to the consideration of the correlation between the initialization and the mean-field potential.

Based on the new version of the model in this work, we study the influence of the slope of symmetry energy, effective mass splitting on the heavy ion collision observables, such as the neutron to proton yield ratios for emitted free nucleons and for coalescence invariant nucleon yield, for $^{124}$Sn+$^{112}$Sn at the beam energy of 200 MeV per nucleon. Our calculations show that the $R(n/p)$ at low kinetic energy region is sensitive to the slope of symmetry energy, and the $R(n/p)$ at high kinetic energy depends on the stiffness of symmetry energy and neutron-proton effective mass splitting. With a given $L$, the inclination of $R(n/p)$ to kinetic energy ($E_k$) can be used to probe the effective mass splitting. On the other hand, if the neutron-proton effective mass splitting is fixed, the values of $R(n/p)$ in the high kinetic energy region increase with $L$, which also implies that the $R(n/p)$ in the high kinetic energy region can be used to probe the symmetry energy above the saturation density. Varying the width of wave packet parameter $\sigma_r$ in a reasonable region does not dramatically change our conclusions. The current calculations show that the analysis of spectra of $R(n/p)$ on the $L$ and $f_I$ parameter space is necessary in future to tightly constrain the symmetry energy.


\section*{Acknowledgements}
This work was partly inspired by the transport code comparison project, and it was supported by the National Natural Science Foundation of China No. 11875323, 11705163, 11790320, 11790323, U1867212, 114222548, and 11961141003, the National Key R\&D Program of China under Grant No. 2018YFA0404404, the Continuous Basic Scientific Research Project (No. WDJC-2019-13, BJ20002501), Innovation Project of Guangxi Graduate Education No. XYCSZ2018060, and the funding of China Institute of Atomic Energy. The work was carried out at National Supercomputer Center in Tianjin, and the calculations were performed on TianHe-1 (A).

\section*{}

\end{document}